%% file: main.tex
\documentclass[conference]{IEEEtran}
\IEEEoverridecommandlockouts

\usepackage{cite}
\usepackage{amsmath,amssymb,amsfonts}
\usepackage{algorithmic}
\usepackage{graphicx}
\usepackage{textcomp}
\usepackage{soul}
\usepackage{xcolor}
\usepackage[acronym]{glossaries}

\input{acronyms}

\newcommand{\rev}[1]{\textcolor{black}{#1}}

\makeatletter 
\newcommand{\linebreakand}{%
  \end{@IEEEauthorhalign}
  \hfill\mbox{}\par
  \mbox{}\hfill\begin{@IEEEauthorhalign}
}
\makeatother 

\newcommand{\bigcell}[2]{\begin{tabular}{@{}#1@{}}#2\end{tabular}}
\def\BibTeX{{\rm B\kern-.05em{\sc i\kern-.025em b}\kern-.08em
    T\kern-.1667em\lower.7ex\hbox{E}\kern-.125emX}}

\usepackage{tikz}
\newcommand{\copyrighttext}{\footnotesize%
979-8-3503-1114-3/23/\$31.00\textcopyright{}2023 IEEE.\\ DOI: 10.1109/VTC2023-Spring57618.2023.10199651 \\}
\newcommand{\copyrightnotice}{%
\begin{tikzpicture}[remember picture,overlay]
\node[xshift=6.2cm,yshift=1.4cm] at (current page.south west) {\parbox{\columnwidth}{\copyrighttext}};
\end{tikzpicture}%
}

\begin{document}

\title{QEVSEC: Quick Electric Vehicle SEcure Charging \\ via Dynamic Wireless Power Transfer}

\author{\IEEEauthorblockN{Tommaso Bianchi}
\IEEEauthorblockA{\textit{Dept. of Mathematics} \\
\textit{University of Padova}\\
Padua, Italy \\
tommaso.bianchi@phd.unipd.it}
\and
\IEEEauthorblockN{Surudhi Asokraj}
\IEEEauthorblockA{\textit{Dept. of Electrical and Computer Engineering} \\
\textit{University of Washington}\\
Seattle, Washington (USA)\\
surudh22@uw.edu}
\and
\IEEEauthorblockN{Alessandro Brighente}
\IEEEauthorblockA{\textit{Dept. of Mathematics} \\
\textit{University of Padova}\\
Padua, Italy \\
alessandro.brighente@unipd.it}
\and

\linebreakand 

\IEEEauthorblockN{Mauro Conti\thanks{Prof. Mauro Conti is also affiliated with University of Washington.}}
\IEEEauthorblockA{\textit{Dept. of Mathematics} \\
\textit{University of Padova}\\
Padua, Italy \\
mauro.conti@unipd.it 
}
\and
\IEEEauthorblockN{Radha Poovendran}
\IEEEauthorblockA{\textit{Dept. of Electrical and Computer Engineering} \\
\textit{University of Washington}\\
Seattle, Washington (USA) \\
rp3@uw.edu}
}

\maketitle

\IEEEpubidadjcol

\copyrightnotice

\input{sections/00_abstract.tex}

\input{sections/01_introduction.tex}

\input{sections/03_models.tex}

\input{sections/04_vulnerabilities.tex}

\input{sections/05_qevsec.tex}

\input{sections/06_analysis.tex}

\input{sections/07_conclusions.tex}

\input{sections/08_ack}

\bibliographystyle{IEEEtran}
\bibliography{references}

\end{document}

%% file: acronyms.tex
\makeglossaries

\newacronym{ev}{EV}{Electric Vehicles}
\newacronym{dwpt}{DWPT}{Dynamic Wireless Power Transfer}
\newacronym{qevsec}{QEVSEC}{Quick Electric Vehicle SEcure Charging}
\newacronym{cp}{CP}{Charging Pads}
\newacronym{cspa}{CSPA}{Charging Service Provider Authority}
\newacronym{ra}{RA}{Registration Authority}
\newacronym{cs}{CS}{Cloud Servers}
\newacronym{fs}{FS}{Fog Servers}
\newacronym{po}{PO}{Pad Owners}
\newacronym{rsu}{RSU}{Road-Side Units}
\newacronym{dsrc}{DSRC}{Dedicated Short-Range Communication}
\newacronym{obu}{OBU}{On-Board Unit}

%% file: sections/00_abstract.tex
\begin{abstract}

\acrfull{dwpt} can be used for on-demand recharging of \acrfull{ev} while driving. However, \acrshort{dwpt} \rev{raises numerous} security and privacy concerns. Recently, researchers \rev{demonstrated} that \acrshort{dwpt} systems are \rev{vulnerable} to adversarial attacks. In an \acrshort{ev} charging scenario, an attacker can \rev{prevent} the authorized customer from charging, \rev{obtain a} free charge by billing a victim user and track a target vehicle. State-of-the-art authentication schemes relying on centralized solutions are either vulnerable to various attacks or have high computational complexity, \rev{making them unsuitable} for a dynamic scenario. In this paper, we propose \acrfull{qevsec}, a novel, secure, and efficient authentication protocol for the dynamic charging of \acrshort{ev}s. Our idea for \acrshort{qevsec} originates from multiple vulnerabilities we found in the state-of-the-art protocol that allows tracking of user activity and \rev{is susceptible to replay attacks. Based on these observations, the proposed protocol solves these issues and achieves lower computational complexity by using only primitive cryptographic operations in a very short message exchange. \acrshort{qevsec} provides scalability and a reduced cost in each iteration, thus lowering the impact on the power needed from the grid.}

\end{abstract}


%% file: sections/01_introduction.tex
\section{Introduction} 

Many countries across the globe are pushing for a transition from fossil-fueled combustion engines to \acrfull{ev}, as gas-powered cars are one of the largest sources of greenhouse gases~\cite{greenhouse}. \rev{As battery-powered} electric motors replace combustion engines in \acrshort{ev}s, \rev{there is} demand for a power supply solution to periodically recharge their battery. 

\acrfull{dwpt}~\cite{hutchinson} is a novel technology that enables charging the \acrshort{ev} while driving. 
Dynamic charging systems provide \acrshort{ev} owners with the flexibility to charge while moving with the help of \acrfull{cp}, i.e., \acrshort{dwpt} base stations embedded under the roads \rev{providing} power to \acrshort{ev}s.  
However, such systems come with novel security and privacy threats, \rev{all of which} are eased by the use of wireless communications for \acrshort{dwpt} and have been \rev{proven} to be critical in \rev{attack} scenarios~\cite{babu2022survey}.

Over the past decade, researchers have developed several privacy-preserving authentication schemes for \acrshort{dwpt} systems. The authentication protocol needs to be lightweight in terms of communication and computation time, secure against different types of attacks, and preserve the privacy of the user. 

Hussain \textit{et al.}~\cite{hussain2015,hussain2017} introduced the idea of \acrshort{cp}s connected to the \acrfull{cspa} and the hash chain-based authentication and revocation of credentials to avoid the fraudulent use of the same. In this protocol, the \acrshort{ev} exchanges multiple messages with the \acrshort{cp}s, which are directly connected to the \acrshort{cspa}. This leads to a large number of interactions between the \acrshort{cp} and \acrshort{cspa} as well as increased utilization of the \acrshort{cp} in the authentication scheme. 

Two other works published by Zhao \textit{et al.}~\cite{zhao} and Rabieh \textit{et al.}~\cite{rabieh} propose authentication schemes for a similar system model. The former uses public-key encryption with a signing and verification scheme provided by a \acrfull{ra}, and a bank in charge of the token provisioning for the charging requests. Each energy segment transmits a constant amount of energy to the \acrshort{ev}. 

The latter scheme~\cite{rabieh} is based on blind signatures, hash chains, and XOR operations. Additionally, the authors address the double-spending attack, in which a malevolent user tries to abuse old credentials to get a free charge. 
 Several other protocols employing multiple entities such as \acrfull{cs}, \acrfull{fs}, \acrfull{po}, and \acrfull{rsu} have been proposed~\cite{babu,roman,gunukula,feng}. However, all these approaches consider a decentralized infrastructure involving multiple mutual authentications between the \acrshort{ev} and various other entities of the \acrshort{dwpt} system. Such exchanges impose higher communication and computational cost on the \acrshort{ev}, hence we focus on centralized systems. 
 \rev{Therefore, we consider a simple model that reduces the attack surface and is likely to be maintained in case of advancement of \acrshort{dwpt} technology.} 

In this paper, we propose \acrfull{qevsec}, a novel, secure, and efficient authentication scheme with enhancements to the vulnerabilities and inefficiencies of the state-of-the-art protocols. Our protocol originates from vulnerabilities we identified in the existing protocol scheme~\cite{hussain2015}, where an attacker can jeopardize the users' location privacy through the charging process and perform a replay attack due to faulty implementation of hash chains. \rev{\acrshort{qevsec} reduces the number of secrets shared between \acrshort{ev} and \acrshort{cspa} to a single one. Thanks to the use of XOR operations, hash functions, and avoiding key-based encryption, we reduce the complexity of the overall authentication process.} Our contributions can be summarized as follows:
\begin{itemize}
    \item \rev{We develop} a new, secure, and efficient protocol, \acrshort{qevsec}, that effectively uses exclusive OR operations, hashing, and hash chains.
    \item \rev{We demonstrate} both via formal analysis and via the Scyther tool, the security of \acrshort{qevsec}.
    \item \rev{Compared} to the state-of-the-art solutions, \acrshort{qevsec} improves the performance during authentication in terms of computation time of around 90\% with a lower linear increment.
\end{itemize}
\par

Our paper is structured as follows: Section II presents the model used in the proposed scheme. Section III presents the drawbacks of our reference state-of-the-art protocol. In Section IV, we describe our solution and improvements, providing the security and performance analysis in Section V. In Section VI we draw our conclusions.

%% file: sections/03_models.tex
\section{System and Adversary Model}

We briefly describe the system model in Section II.A, and the adversary model in Section II.B.

\subsection{System Model}
\begin{figure}[!h]
    \begin{center}
        \includegraphics[width=\columnwidth]{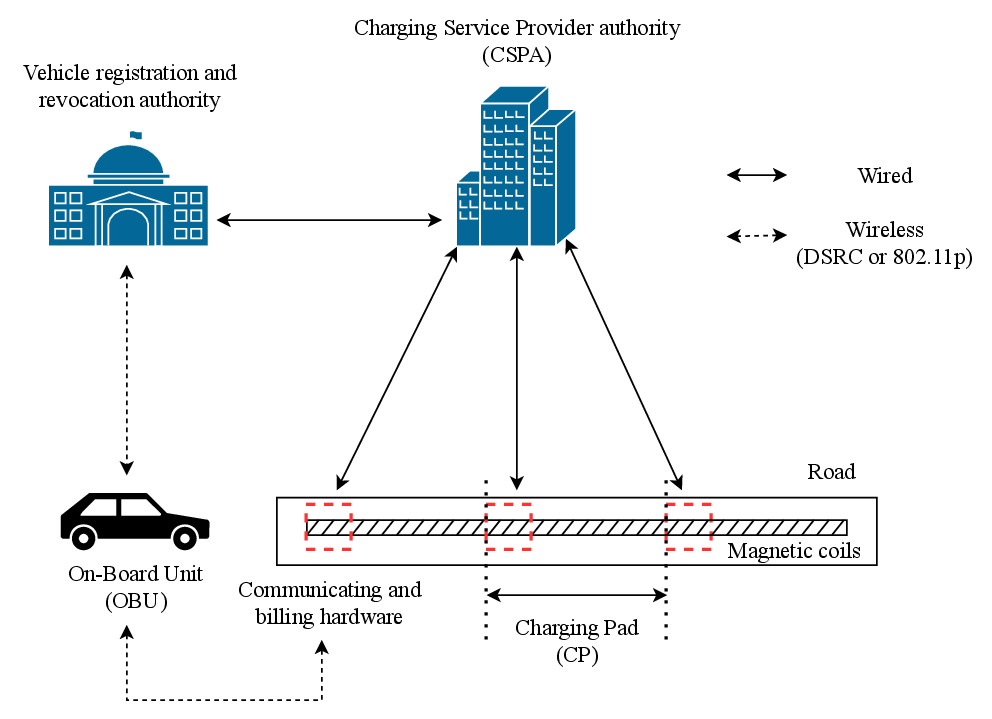}
        \caption{Our considered system model\rev{, that involves the \acrshort{obu}, \acrshort{ra}, \acrshort{cspa}, and \acrshort{cp}s. The beginning part of these last is a short segment dedicated to communication and computation hardware (indicated with red boxes) that performs all the cryptographic operations.}}
        \label{fig:model_scheme}
    \end{center}
\end{figure}

We consider four different entities: \acrshort{ev}, \acrshort{cp}, \acrshort{cspa}, and \acrshort{ra}. In our model, the \acrshort{cspa} is directly connected via a wired \rev{connection} to the \acrshort{ra} and all the \acrshort{cp}s. \rev{The schematic model of the network architecture has been presented in Fig.}~\ref{fig:model_scheme}. \acrshort{ev} and \acrshort{cp}s can directly communicate and authenticate \rev{each} another, whereas the \acrshort{cspa} can be involved or not, depending on the protocol scheme. The \acrshort{ev} contains an \acrfull{obu} that manages the cryptographic operations and securely stores the EV's parameters, including the \acrshort{ev} identity and its pseudonyms. We consider the \acrshort{obu} to be secure and tamper-proof. The \acrshort{ra} is responsible for publishing the parameters for the encryption scheme and generating the pseudonyms for the \acrshort{obu} at the time of registration. The communication between the different facilities and the \acrshort{obu} may either happen through \acrfull{dsrc} or other wireless communication protocols.

\subsection{Adversary Model}
    
Our system considers both the \acrshort{ev} and the \acrshort{cspa} as malicious and not trustworthy. The adversary can compose, replay, intercept, and forge messages, but \rev{they} cannot decipher \rev{the message} without the correct cryptographic keys. The goal of the attacker is to infer the private key between the two entities to \rev{obtain} all the parameters of the charging process or to intercept and replay packets to trigger some action by \acrshort{ev} or \acrshort{cspa}. Successively, \rev{they} can launch attacks to get a free charge or identify the vehicle, mining the privacy of the customer. 
Only the \acrshort{ra} is trustworthy and knows the true identity of the \acrshort{ev}s, which is never revealed during the authentication process. Because of the symmetric key encryption, the wired communication network is considered secure; however, an adversary can connect to the network and sniff the traffic. If the protocol is poorly designed, an adversary may infer information for further attacks, such as tracing or replay attacks. This adversary representation is formulated in the Dolev-Yao model~\cite{dolev-yao}.

%% file: sections/04_vulnerabilities.tex
 \section{Vulnerabilities and Attacks}
In this section, the problems identified in the \rev{reference} protocol \cite{hussain2015} \rev{is described.} 
The first challenge is related to the use of the same identifier throughout multiple authentication runs, which \rev{can expose the customer's identity and allow tracking of it in different charging processes.} This poses a severe threat to the \acrshort{dwpt} system privacy, as it will enable tracing the location of the customer by looking at her interaction with the system.

The second research challenge is related to the hash-chain approach $h^n(x) \rightarrow h(h^{n-1}(x))$ used for \acrshort{obu}-\acrshort{cp} authentication, and particularly how the \acrshort{cspa} updates the value for the next expected hash chain value $x$ for a hash function $h$.
Instead of storing the current value received from the \acrshort{obu} (i.e., $h^{n-1}(x)$), the \acrshort{cspa} in \cite{hussain2015} stores the hash value that is already in memory, resulting in the same hash chain parameter received at the beginning of the protocol being stored at each iteration:

\begin{equation}
h(h^{n-1}(x)) = h^n(x).
\end{equation}

This \rev{behavior} has two consequences. First, the \acrshort{obu} currently participating in the protocol exchange is unable to authenticate itself further. This occurs because the hash value that the \acrshort{obu} sends (e.g., $h^{n-2}(x)$) and the value that \acrshort{cp} or \acrshort{cspa} expects (e.g., $h^{n-1}(x)$) are different. Following the initial successful authentication, the \acrshort{cspa} repeatedly waits for the same value, resulting in an error.
 Second, the \acrshort{obu} could send the same $h^{n-1}()$ value for authentication indefinitely, and an attacker could eavesdrop on the packet and then pose as the authenticated vehicle. Thus, this can result in a successful free-riding attack (i.e., get a free charge by billing another customer) through a replay attack.

To mitigate this vulnerability, it is sufficient to store the most recent value received from \acrshort{obu}, i.e.,

\begin{equation}
h^i(PS^i_{OBU}) \rightarrow h^{i-1}(PS^i_{OBU}).
\end{equation}

%% file: sections/05_qevsec.tex
\section{\acrshort{qevsec}} 

In this section, we present our protocol \acrshort{qevsec}. Due to space constraints, we do not reintroduce the protocol in \cite{hussain2015}. We point the reader to \cite{hussain2015} to grasp the differences between the original protocol and our proposal.

The first step is to provide a way to verify the veracity of the \acrshort{obu} registration later in the scheme. We allow it by storing a copy of the \acrshort{ra} database of vehicle pseudonyms with the pairs $(X_{OBU}, z_i)$ at \acrshort{cspa}, with $z_i$ being a different random number associated with each $X_{OBU}$. We define the value of $X_{OBU}$ as $h_2(PS^i_{OBU})$, where $h_2$ is a collision-free hash function provided by \acrshort{ra} at scheme initialization, and $PS^i_{OBU}$ is the pseudonym that \acrshort{cspa} generates for the vehicle. In this way, we generate secrets between \acrshort{cspa} and \acrshort{obu} without revealing the mapping between pseudonyms.  The random values are distributed to \acrshort{obu}s along with the corresponding $PS^i_{OBU}$, in order to provide a common secret at the beginning of the authentication. Each vehicle has different $PS^i_{OBU}$ in order to use them for different charging sessions. In fact, a vehicle never utilizes the same $PS^i$ value more than once.

Following the initialization phase, in the first step \acrshort{ev} sends to \acrshort{cspa} the message 

\begin{equation}
OBU \rightarrow CSPA: m_1 = (X^i_{OBU}).
\end{equation}

\acrshort{cspa}, after receiving the message, constructs the three parameters $H_1 = h(s \parallel X^i_{OBU})$, $H_2 = h(H_1) \oplus z_i$, and $H_3 = MSK_i \oplus H_1$, where $s$ is a secret only known by \acrshort{ra}, and $MSK_i$ is $i$-th \acrshort{ra}'s secret key to use with each different vehicle. $H_1$ is stored as a security parameter, and \acrshort{cspa} sends only the other two values along with the hash function and a check parameter $check = h(X^i_{OBU} \oplus z_i)$ in
\begin{equation}
CSPA \rightarrow OBU: m_2 = (h(), H_2, H_3, check). 
\end{equation}
In this way, the \acrshort{obu} can verify the knowledge of the correct $z_i$ from \acrshort{cspa}. To avoid overhead message authentication between \acrshort{ev} and \acrshort{cp} that generates expensive operations in constrained devices, all the $m_1$ to $m_4$ message exchange occurs between the \acrshort{obu} and the \acrshort{cspa}. As a result, \acrshort{obu} mutually authenticates with the system's first level (the \acrshort{cspa}), and then it uses a hash chain to authenticate with the \acrshort{cp}s. Therefore, we continue with the following messages to verify the common secret:
\begin{equation}
OBU \rightarrow CSPA: m_3 = (c_1, c_2, c_3, c_4, H_3),
\end{equation}
where 
\begin{gather}
    c_1 = h(H_2) \oplus PS^i_{OBU}, \, c_2 = h(h(PS^i_{OBU}) \parallel H_3),\\
    c_3 = r_{OBU} \oplus PS^i_{OBU}, \, c_4 = h^n(N_{OBU} \parallel PS^i_{OBU}) \oplus z_i.
\end{gather}

\begin{figure}[!t]
    \begin{center}
        \includegraphics[width=0.95\columnwidth]{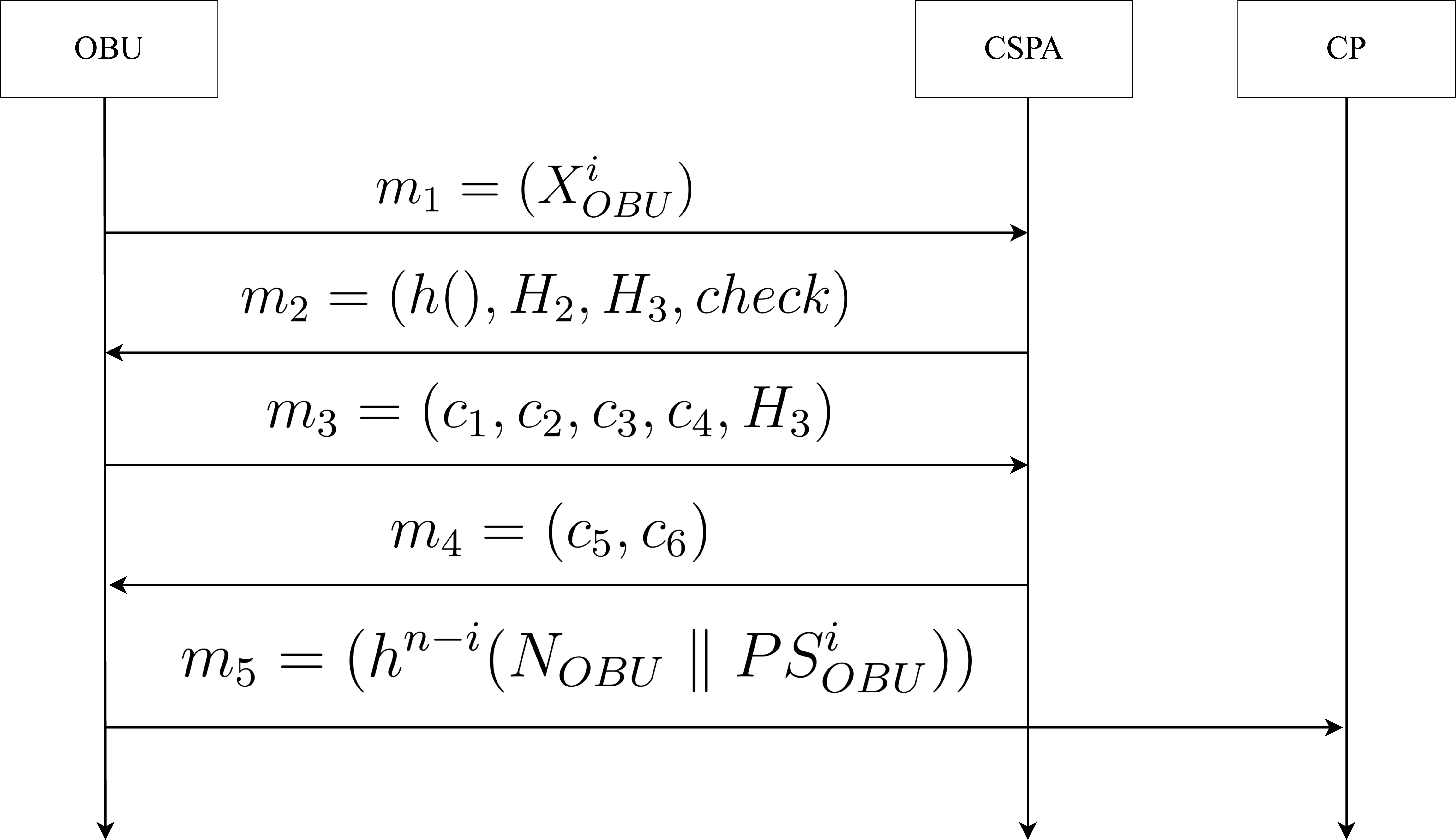}
        \caption{\rev{Diagram scheme of \acrshort{qevsec} protocol. After $m_5$, the \acrshort{obu} sends to the next \acrshort{cp} the subsequent value in the hash-chain for authentication.}
        }
        \label{fig:revised_scheme}
    \end{center}
\end{figure}

In particular, in $c_1$ the value of $h(H_2)$ \rev{is hidden} with the exclusive OR so that \acrshort{cspa} can verify it using the right pseudonym. We use $c_2 = h(h(PS^i_{OBU}) \parallel H_3)$ to check if the received message is correct and from the right source. In $c_3$, 
the exclusive OR between $PS^i_{OBU}$ and $r_{OBU}$ \rev{is utilized}, with $r$ being a random number. \acrshort{obu} generates a nonce $N_{OBU}$  in order to generate the hash chain as $h^n(N_{OBU} \parallel PS^i_{OBU})$, sent hidden by the exclusive OR operation with the common secret in $c_4$.
In this way, the values of $H_2$ (hidden in the XOR operation) and $H_3$ (constructed by values known only by \acrshort{cspa}) can be sent in clear at the beginning of the protocol without leaking information that can be used for a Man-In-The-Middle attack. \acrshort{cspa} can use the value of $H_1$ extracted from $H_3$ as in the original work \cite{hussain2015}, and consequently extract $PS^i_{OBU}$ from $c_1$, $r_{OBU}$ from $c_3$ and compare the results to $c_2$. The hash chain extracted from $c_4$ is used as an authentication parameter between \acrshort{obu} and each successive \acrshort{cp} during the charging process. In the last step of \acrshort{ev}-\acrshort{cspa} authentication, similar to $r_{OBU}$, \acrshort{cspa} generates a nonce for the run of the protocol, $r_{CSPA}$. \acrshort{cspa} sends the following to \acrshort{obu}:
\begin{equation}
CSPA \rightarrow OBU: m_4 = (c_5, c_6).
\end{equation}
where
 \begin{gather}
    c_5 = P \oplus r_{CSPA}, \\
    c_6 = g^{(P \oplus (r_{CSPA} - n))}, \\
    P = h(r_{OBU} \parallel PS^i_{OBU}).
\end{gather}

From this message, \acrshort{obu} can extract the value of $r_{CSPA}$ and check the result of the exponentiation against $c_4$. This approach necessitates the publication of the parameters $g$ and $n$, which are constant during the entire protocol. In the next phase, the \acrshort{cspa} sends the hash chain value provided by \acrshort{obu} to the \acrshort{cp}s to authenticate and begin the charging process as in the \textit{hash chain-based authentication} proposed in \cite{hussain2015}. \rev{Fig. \ref{fig:revised_scheme} shows our protocol diagram with the message exchange.}

%% file: sections/06_analysis.tex
\section{Security and Performance Analysis}
In this section, we first prove the security of \acrshort{qevsec} using BAN Logic~\cite{ban} and Scyther tool and then compare it with the state-of-the-art in terms of communication costs. 

BAN logic uses the concept of \textit{belief}, where the entities involved trust the state of the protocol in terms of \textit{freshness} and \textit{shared secrets}. The former indicates messages sent and received with a nonce or new terms that implicate the freshness of a later packet. The latter is based on two or more parties sharing a valid secret that, if not leaked, indicates that one of the two trusted parties sent a message with this term. 
The final result of the procedure is a state where both entities involved trust the messages and secrets inside them, without leaking information to third parties. Table~\ref{table:ban_constructs} describes the constructs \rev{used} to prove the security and usability of \acrshort{qevsec}.

\input{tables/ban_logic_table}

We start by showing that the first two messages $m_1$ and $m_2$ allow for a secure exchange, without letting an attacker infer data or get an advantage in replaying the packets. When sending $m_1$
${\rm CSPA}|\# X^i_{OBU}$,
i.e., the provider recognizes $X^i_{OBU}$ as a fresh parameter, assuming that it was never used by \acrshort{ev} or revoked.
After $m_2$ is sent, we can affirm
$EV<| (H_2, H_3, check)$,
and the expressions follow the rules, in order, \textit{Shared secret rule}, \textit{Freshness rule}, and \textit{\rev{nonce} verification rule}, i.e.,
\begin{gather}
\frac{EV|= {\rm CSPA} = z_{i} = EV, EV<| check}{EV |= {\rm CSPA} |\sim (H_2, H_3, check)}, \\
\frac{EV |= \#(check)}{EV |= \#(H_2, H_3, check) }, \\
\frac{EV |= \#(H_2), EV |= {\rm CSPA} |\sim (H_2, H_3, check)}{EV |= {\rm CSPA} |= (H_2, H_3, check)}.
\end{gather}
 From now on, the three rules are used in the same sequence for messages $m_3$ and $m_4$. 
 
 Following the above statements, \acrshort{ev}, recognizing the freshness of the message received by \acrshort{cspa}, \textit{believes} the other entity with the packet received $EV |= {\rm CSPA} |\sim (H_2, H_3, check)$,
$EV |= {\rm CSPA} |= (H_2, H_3, check)$.
Similarly, for the parameters in message $m_3$, ${\rm CSPA} |= EV | \sim (c1, c'_3, c'_4, c'_5, H_3)$,
i.e., \acrshort{cspa} believes in the packet sent by \acrshort{ev} thanks to the extraction of $H_1$ from $H_3$ and the retrieval of $PS^i_{ \rm OBU}$:
\begin{gather}
\frac{{\rm CSPA}|= EV = h(h(PS^i_{ \rm OBU})) = {\rm CSPA}, {\rm CSPA} <| PS^i_{\rm OBU}}{{\rm CSPA} |= EV |\sim (c1, c'_3, c'_4, c'_5, H_3)},\\
\frac{{\rm CSPA} |= \#(PS^i_{ \rm OBU})}{{\rm CSPA} |= \#(c1, c'_3, c'_4, c'_5, H_3)}, \\
\frac{{\rm CSPA} |= \#(PS^i_{\rm OBU}), {\rm CSPA} |= EV |\sim (c1, c'_3, c'_4, c'_5, H_3)}{{\rm CSPA} |= EV |= (c1, c'_3, c'_4, c'_5, H_3)}.
\end{gather}
As before, we show that \acrshort{cspa} believes the parameters retrieved by this last message, considering that it trusts the secrecy of $z_i$, i.e., ${\rm CSPA} |= EV |\sim PS^i_{ \rm OBU}, {\rm CSPA} |= EV |= PS^i_{\rm OBU}, {\rm CSPA} |= EV = PS^i_{ \rm OBU} = {\rm CSPA}$.
Generally, the entire $m_3$ is trusted, including the nonce $r_{OBU}$. 
With the last message, we can conclude that
\begin{gather}
\frac{EV |= {\rm CSPA} = P' = EV, EV <| (c'_6, c'_7)}{EV |= {\rm CSPA} |\sim (c'_6, c'_7)},\\
\frac{EV |= \#P'}{EV |= \#(c'_6, c'_7)}, \\
\frac{EV |= \#P', EV |= {\rm CSPA} |\sim (c'_6, c'_7)}{EV |= {\rm CSPA} |= (c'_6, c'_7)}.
\end{gather}

\begin{figure}[h!]
    \centering
    \includegraphics[width=0.85\columnwidth]{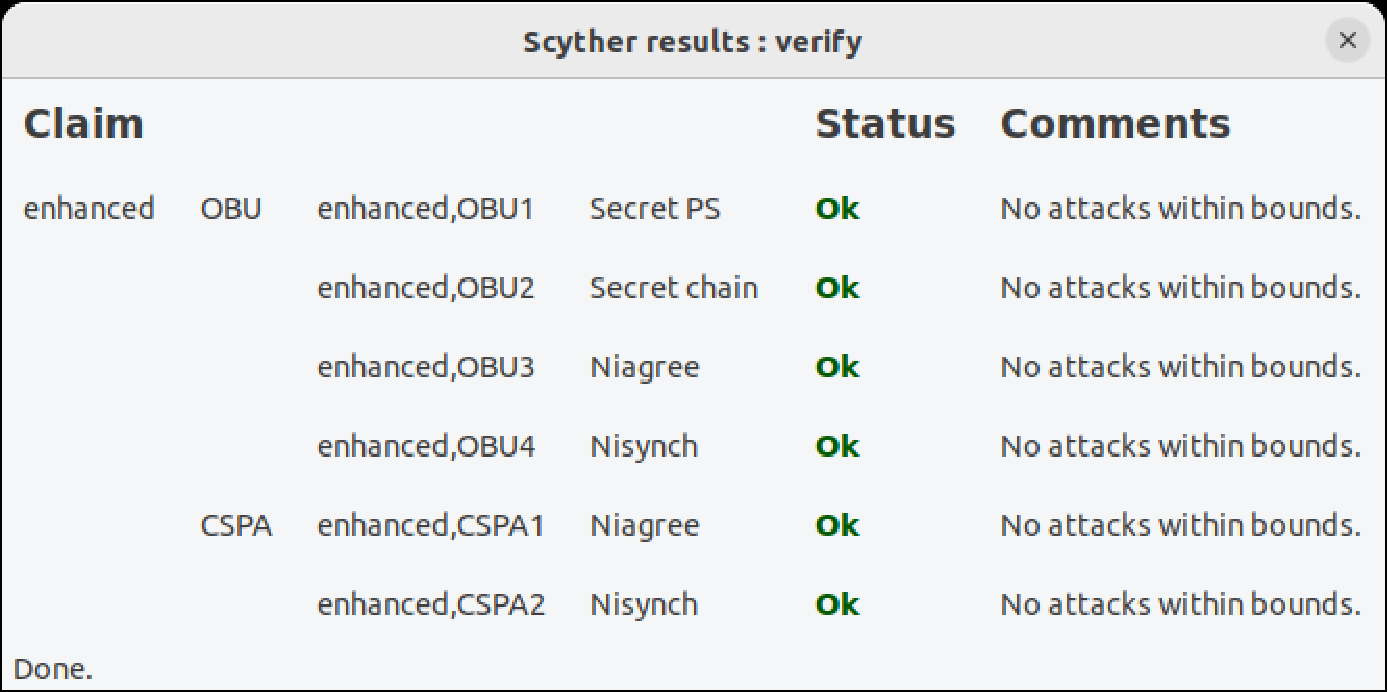}
    \caption{Scyther tool results for secrecy of $PS^i_{ \rm OBU}$ and the hash-chain head. \rev{The outcomes prove the impossibility of inferring and stealing private information during different protocol runs.}
    }
    \label{fig:scyther}
\end{figure}

\begin{table*}[!h]
\begin{center}
    \caption{Computational cost comparison between QEVSEC and the most related state-of-the-art approaches.}
    \begin{tabular}{|p{2cm}|p{3.6cm}|p{3.25cm}|p{3.25cm}|p{3cm}|}
        \hline
        & \textbf{Auth. \acrshort{obu}} [ms] & \textbf{Auth. \acrshort{cp}/\acrshort{cspa}} [ms] & \textbf{Hash-chain} [ms] & \textbf{Total time} [ms]\\
        \hline
        \textbf{Hussain \textit{et al.}} \cite{hussain2015} & 6$T_{h}$ + 6$T_{XOR}$ = 1.62  &  7$T_{h}$ + 6$T_{XOR}$ = 1.89 & \textemdash & $3.51 \times n$ \\
        \hline
        \textbf{Rabieh \textit{et al.}} \cite{rabieh} & 4$T_{exp}$ + 4$T_{ecm}$ + 2$T_{ver}$ + $T_{sig}$ + $T_{h}$ = 10.01  & 2$T_{pair}$ + 4$T_{ecm}$ + 4$T_{exp}$ + $T_{sig}$ + $T_{ver}$ = 10.01  & $n \times 0.27$ & $20.02 +  T_{Hash-chain}$  \\
        \hline
        \textbf{Zhao \textit{et al.}} \cite{zhao} & 2$T_{sig}$ + 2$T_{ver}$ + $T_{h}$ = 5.15  & $T_{sig}$ + 2$T_{ver}$ = 3.89 & $ n \times 0.27 + n \times (T_{sig} + T_{ver})$ & $9.04 + T_{Hash-chain}$ \\
        \hline
        \textbf{\acrshort{qevsec}} &\textbf{6$\mathbf{T_h}$ + 5$\mathbf{T_{XOR}}$ + $\mathbf{T_{exp}}$ = 1.73 } & 
        \textbf{8$\mathbf{T_{h}}$ + 5$\mathbf{T_{XOR}}$ + $\mathbf{T_{exp}}$ = 2.27 } & \textbf{$\mathbf{n \times 0.27}$} & \textbf{$\mathbf{4.00 + }$  $\mathbf{T_{Hash-chain}}$ } \\
        \hline
    \end{tabular}
    \label{table:comparison}
\end{center}
\end{table*}

\begin{table}[!h]
    \begin{center}
        \caption{Cryptographic primitive execution time.}
        \begin{tabular}{|p{1.75cm}|p{2.75cm}|} 
            \hline  
            \textbf{Primitive} & \textbf{Average Time (ms)} \\
            \hline
            $T_{exp}$ & 0.110 \\ 
            \hline
            $T_{pair}$ & 0.884 \\ 
            \hline
            $T_{h}$ & 0.27  \\ 
            \hline
            $T_{ecm}$ & 1.352 \\ 
            \hline
            $T_{ver}$  & 1.449 \\ 
            \hline
            $T_{sig}$  & 0.992 \\ 
            \hline
        \end{tabular}
        \label{table:computational_cost}
    \end{center}
\end{table}

Finally, \acrshort{ev} recognizes the validity of \acrshort{cspa}, with the secrecy and freshness of the different parameters involved during the authentication, including nonce $r_{{\rm CSPA}}$, i.e., $EV |= {\rm CSPA} |\sim (c'_6, c'_7), EV |= {\rm CSPA} = P' = EV, EV |= {\rm CSPA} |= (c'_6, c'_7)$.
This concludes the proof for the impossibility of performing a replay attack. 

To further prove the security of \acrshort{qevsec}, we test it with Scyther tool~\cite{scyther}, a program used to inspect a cryptographic protocol in order to find possible attacks. Fig.~\ref{fig:scyther} shows the results of the tool, proving the security of the most important parameters in the scheme. \acrshort{qevsec} is secure in different consecutive runs, maintaining the secrecy of the different parameters for the \acrshort{ev}.

\begin{figure}[!h]
    \centering
    \includegraphics[width=0.99\columnwidth]{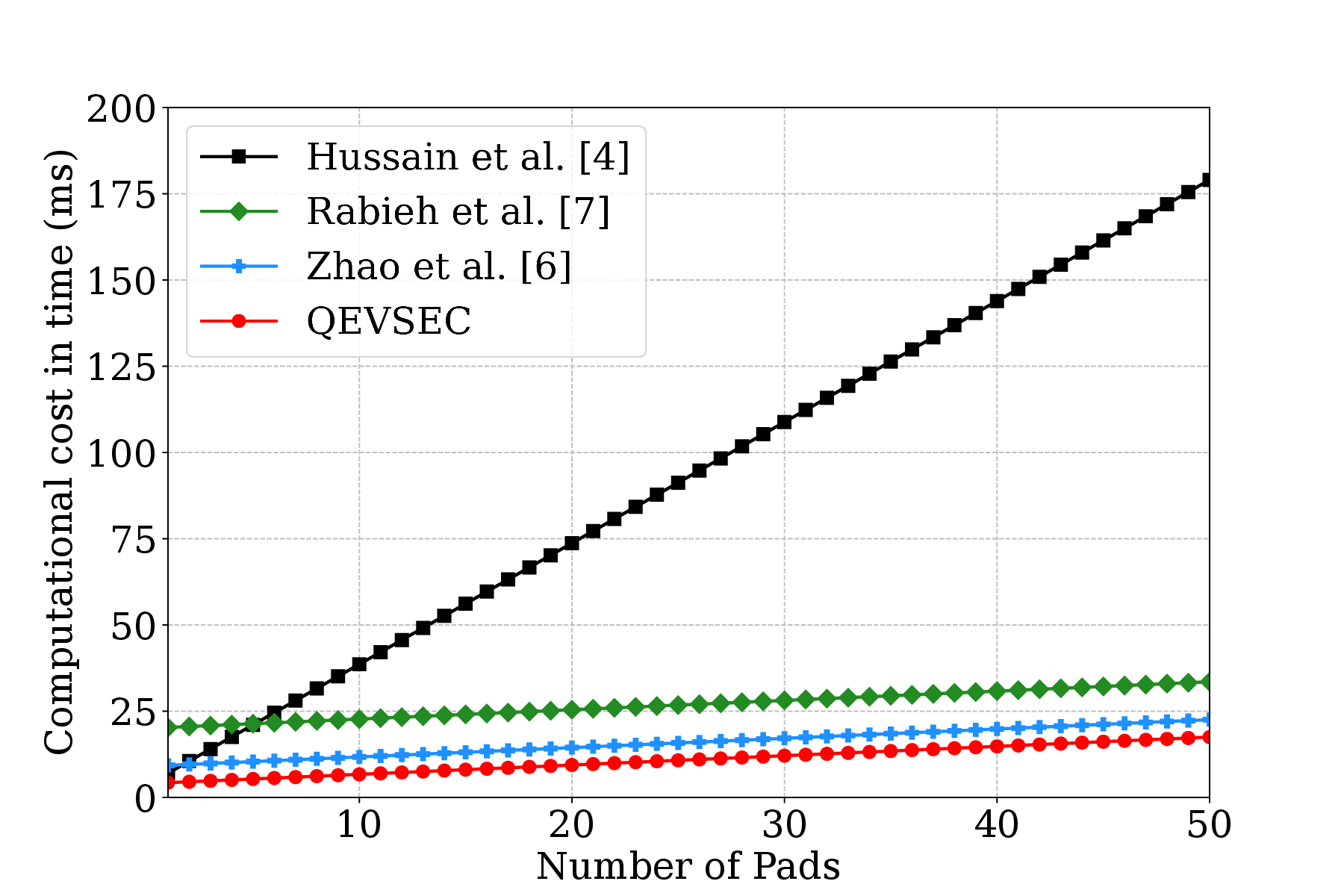}
    \caption{Computational time against the number of charging pads. 
    }
    \label{fig:comp_cost_comparison}
\end{figure}

\rev{To conclude the analysis, a comparison is provided by counting} the number of operations in the authentication steps between \acrshort{obu}-\acrshort{cspa} and \acrshort{obu}-\acrshort{cp}.  Table~\ref{table:computational_cost} reports the time taken by the operations in the modified scheme\rev{, computed using a simulation of the primitives in Python, using Charm-Crypto Library~\cite{charm}. In order, $T_{exp}$ and $T_{pair}$ are the time for exponentiation and pairing, respectively. $T_{h}$ is the hash computation time, while $T_{ecm}$ is the cost for the elliptic curve multiplication. Finally, $T_{ver}$ and $T_{sig}$ are the time for signing end verification in \cite{rabieh} and \cite{zhao}. These time values are used to compare the different protocols with a common basis for the computational cost of the primitive operations.}
The \textit{exclusive or} operation consumes little time, and hence, \rev{it is excluded} in the results. In Table~\ref{table:comparison}, \rev{ we report the cost that each step of authentication takes in time, using "\textemdash" when no messages are exchanged in that phase. The total time taken by \cite{hussain2015} for a single \acrshort{cp} is lower by 0.5 ms with respect to \acrshort{qevsec}, but this value has to be multiplied for each pad used during the charging process. Considering the integration of the hash chain at the CP level, we achieve better performance after the first constant part of the protocol. Following that, instead of the entire computation cost for the authentication process, as in \cite{hussain2015}, we need only a message containing a hash, thus remarkably reducing the overall cost. Our protocol has minimal overhead stemming from the pre-authentication message exchange required to eliminate the usage of any parameter that can be used to track user activity.} 
Rabieh \textit{et al.}~\cite{rabieh} comprises a digital signature, while in Zhao \textit{et al.} scheme~\cite{zhao}, also a digital signature is requested for each plate, making the two schemes computationally heavier than \acrshort{qevsec}. In \rev{Fig.} \ref{fig:comp_cost_comparison}, we show the results for the computation time with respect to the number of pads used in the charging process. Our protocol has  a lower setup time and the hash chain allows a linearly incremental time that maintains the total time lower than the other protocols.

%% file: tables/ban_logic_table.tex
\begin{table}[!th]
    \caption{BAN constructs for the proof.}
    \label{table:ban_constructs}
    \begin{center}
        \begin{tabular}{|p{2cm}|p{6cm}|}
            \hline
            \textbf{Notation} & \textbf{Description} \\
            \hline 
            $P |= X$ & \bigcell{l}{P believes X so P thinks that X \\ is true.} \\ \hline
            $P <| X$ & P sees message X. \\ \hline
            $P |\sim X$ & P once said X. \\ \hline
            $\#(X)$ & X is fresh. \\ \hline
            
            $P=X=Q$ & X is a secret known only by P and Q \\ \hline
            
            Shared Key Rule & $\frac{P |= Q <- K -> P, P <| \{x\}K}{P |= Q |\sim X}$, If P believes that K is a good K and P sees X encrypted with K, then P believes that Q once said X.  \\ \hline
            
            Nonce Verification Rule & $\frac{P |= \#(X), P |= Q |\sim X}{P |= Q |= X}$, the only formula in order to promote $|\sim$  to $|=$, says that P believes X to be  recent, and Q said X, then P believes  that Q believes X. \\ \hline
            
            Freshness Rule & $\frac{P |= \#(X)}{P |= \#(X, Y)}$, if part of the formula is fresh, the entire formula is believed to be fresh. \\ \hline
           
        \end{tabular}
    \end{center}
\end{table}

%% file: sections/07_conclusions.tex
\section{Conclusion}
 The vulnerabilities we identified in the state-of-the-art allow the adversary to attack the charging infrastructure or \acrshort{ev} by eavesdropping, intercepting, and tampering with the exchanged messages. We propose \acrshort{qevsec}, an enhanced, lightweight, and secure authentication protocol that improves system security by eliminating threats while lowering the computational costs of the system. \acrshort{qevsec} protects the \acrshort{ev} from adversarial attacks including but not limited to replay and denial-of-service attacks. Furthermore, it provides scalability with respect to the number of pads. We also proved the security of our scheme against replay attacks and its secrecy via both formal analysis and an automated tool. Our comparison with other state-of-the-art approaches shows that \acrshort{qevsec} is the best-performing solution in terms of computational cost.

%% file: sections/08_ack.tex
\section*{Acknowledgment}

This work was supported by the Office of Naval Research via Grant N00014-20-1-2636 and by the European Commission under the Horizon Europe Programme, as part of the project LAZARUS (https://lazarus-he.eu/) (Grant Agreement no. 101070303).   \par
The content of this article does not reflect the official opinion of the European Union. Responsibility for the information and views expressed therein lies entirely with the authors.